\documentclass[aps,showpacs,floatfix,prl,twocolumn]{revtex4}
\usepackage{amssymb,amsfonts,amsmath,bm,graphicx}
\usepackage{hyperref}

\makeatother

\begin{document}

\title{Effective chiral restoration in the $\rho'$-meson in lattice QCD}

\author{L.~Ya.\ Glozman, C.\! B.\ Lang, and Markus Limmer}

\affiliation{Institut f\"ur Physik, FB Theoretische Physik,
Universit\"at Graz, A-8010 Graz, Austria}

\begin{abstract}
In simulations with dynamical quarks it has been established that the ground
state $\rho$ in the infrared is a  strong mixture of the two chiral
representations $(0,1)+(1,0)$ and $(1/2,1/2)_b$. Its angular momentum content
is approximately the $^3S_1$ partial wave which is  consistent with the quark
model. Effective chiral restoration in an excited $\rho$-meson would require
that in the infrared this meson couples predominantly to one of the two 
representations. The variational method allows one to study the mixing of 
interpolators with different chiral transformation properties in the
non-perturbatively determined  excited state at different resolution scales. We
present results for the  first excited state of the $\rho$-meson using
simulations with $n_f=2$ dynamical quarks. We point out, that in the infrared a
leading contribution to  $\rho'= \rho(1450)$ comes from $(1/2,1/2)_b$, in
contrast to the $\rho$. Its approximate chiral partner would be a $h_1(1380)$
state. The $\rho'$ wave function contains a significant contribution of the 
$^3D_1$ wave which is not consistent with the quark model prediction.
\end{abstract}



\maketitle

\paragraph{1.~Introduction.}

A central question for QCD is how both, confinement and chiral symmetry
breaking, are interrelated and influence the mass generation of hadrons. Chiral
symmetry is  dynamically broken in the QCD vacuum and this phenomenon is the
most important for the mass origin of the ground state hadrons. It is believed,
that a coupling of valence quarks with the quark condensate of the vacuum is
responsible for the large constituent mass of quarks at low momenta. This large mass
makes  the problem effectively non-relativistic and the ground state $\rho$ is a
$^3S_1$ state according to the quark model language \cite{KO,CL,IS,PDG}.
Traditionally the excitation spectrum is also described with the quark model
and, e.g., the first excited state of the $\rho$-meson, $\rho(1450)$, is
believed to be the first radial excitation of the ground  state, i.e., the
$^3S_1$ state \cite{IS,PDG}.

The empirical  spectrum of the highly excited hadrons in the light quark sector
exhibits  patterns of parity doublets. It has been suggested that these patterns
signal effective restoration of both $SU(2)_L \times SU(2)_R$ and $U(1)_A$
symmetries in excited hadrons \cite{CG0,CG1} (for a review see \cite{CG2}). This
conjecture would imply that the mass generation mechanism in highly excited
hadrons is very different and the quark condensate of the vacuum  is of little
importance. It would also imply that the constituent quark model language is not
adequate for highly excited states.

Such an interpretation of the spectroscopic patterns is not unique and it is
possible to imagine some alternative explanations of existing symmetry patterns
\cite{JPS1,SV,A,K}. Effective chiral and $U(1)_A$ restorations have a strong
predictive power. Some of the chiral partners for existing highly excited states
are  missing and one obvious way to confirm or reject effective chiral and
$U(1)_A$ restorations is to find experimentally missing chiral partners or
establish their absence \cite{GS}. This is a difficult experimental
task, though. Another empirical approach to answer this question is to find
alternative experimental signatures that would correlate with the spectroscopic
patterns. Indeed, the effective chiral restoration would require that the states
with approximate chiral symmetry must almost  decouple from pions and their
diagonal axial coupling constants must be small \cite{CG3,JPS2}. It is
difficult, if not impossible, to measure experimentally diagonal axial coupling
constants for highly excited hadrons. The effective chiral restoration
also requires that the states which are assumed to be in approximate chiral
multiplets, have a  small strong decay coupling constants into the ground states
and a pion. The analysis of empirical decays of excited nucleons shows that
indeed all those excited nucleons that are approximate parity doublets have a
very small decay coupling constant  \cite{G1}. There are no other experimental
tools that would help us to clarify this important question beyond reasonable
doubts. To resolve the issue one needs direct information about the hadron
structure, which can be supplied in ab initio lattice simulations.

A first attempt to address the problem on the lattice was Ref.~\cite{DG}. The
low-lying states are dominated by the near zero modes of the Dirac operator,
which are directly related to the quark condensate of the vacuum. The role of
these modes at  small  Euclidean times of the two-point correlation function is
insignificant. This part of the correlator is dominated by the highly
excited hadrons. This observation is consistent with effective chiral
restoration but not sufficiently clear. A conclusive evidence  would  require to
extract the high-lying chiral partners and see what the role of the near zero
modes for the splitting of these states is. This is a difficult task, however.
The other way is to measure the axial coupling constants of excited states
\cite{TK}. For that purpose one would need reliable plateaus in two- and
three-point functions for excited states near the chiral limit, which is not
easy.

Here we suggest an alternative approach to the problem. Using the variational
method \cite{Mi85} and a set of interpolators that scan a complete set of chiral
representations for a given meson, we can study couplings of different
interpolators to a given meson at different resolution scales \cite{GLL}. This
method has successfully been applied to the ground state $\rho$-meson and the
analysis has revealed that the ground state in the infrared is a strong mixture
of the two possible representations  $(0,1)+(1,0)$ and $(1/2,1/2)_b$ of $SU(2)_L
\times SU(2)_R$. Chiral symmetry is strongly broken in the state and this state
is approximately a $^3S_1$ partial wave which is in agreement with the quark
model language. Here we use this method for the first excited state of the
$\rho$-meson, $\rho(1450)$. We are able to present for the first time  a clear
evidence that in the infrared the structure of this state is different from the
quark model prediction ($^3S_1$): it belongs predominantly to the $(1/2,1/2)_b$
representation, which indicates a smooth onset of effective chiral restoration.

\paragraph{2.~Elements of the formalism.}

There exist two different local operators with the $\rho$-meson quantum numbers
$I, J^{PC} = 1, 1^{--}$, the vector current, $\bar{q}\gamma^i{\bm\tau}q$, and
the pseudotensor ``current", $\bar{q}\sigma^{0i}{\bm\tau}q$. It is well
established, both in quenched and dynamical
(see, e.g., \cite{BeLuMe03,BrBuGa03,BuGaGl06,AlAnAo08,GLL,Engel:2010my}) lattice simulations, that both the ground state
$\rho$-meson as well as its first excited state can be created from the vacuum
by either of these operators. These two operators have distinct chiral
transformation properties with respect to $SU(2)_L \times SU(2)_R$
\cite{CoJi97,CG1,CG2}. The vector current belongs to the $(0,1)+(1,0)$
representation and its chiral partner is the axial-vector current
$\bar{q}\gamma^i \gamma^5 {\bm\tau}q$, which creates from the vacuum the
axial-vector meson $a_1$ ($I, J^{PC} = 1, 1^{++}$). The pseudotensor
interpolator transforms as $(1/2,1/2)_b$ and its chiral partner is the operator
$\epsilon^{ijk}\bar{q}\sigma^{jk} q$ that creates from the vacuum the
$h_1$-meson ($I,J^{PC} = 0, 1^{+-}$).

Assuming that chiral symmetry is not broken both explicitly and spontaneously,
there would be two different groups of $\rho$-mesons. The first group would belong
to the $(0,1)+(1,0)$ representation, would couple exclusively to the vector
current (or to the non-local operator with the same chiral transformation
properties) and each member would be mass degenerate with its
corresponding axial-vector partner $a_1$.
The $\rho$-mesons of the second group would transform as $(1/2,1/2)_b$, could be
created from the vacuum only by the pseudotensor operator (or by the non-local
operator with the same chiral transformation properties) and would be
systematically degenerate with their corresponding partners $h_1$. 
A total amount of  $\rho$-mesons
in the spectrum would coincide with the combined amount of $a_1$- and
$h_1$-mesons.

Chiral symmetry breaking in the vacuum implies that the $\rho$-mesons are
mixtures of both the $(0,1)+(1,0)$ and $(1/2,1/2)_b$ representations and can be
created by both  the vector and the pseudotensor interpolators.  Effective
chiral restoration in highly excited $\rho$-mesons would require that some of them
predominantly couple to the vector interpolator, and the other ones would couple
predominantly to the pseudotensor operator. Asymptotically each of these mesons
would belong entirely to one of the two representations, $(0,1)+(1,0)$ or
$(1/2,1/2)_b$, and could not be created by the operator that transforms
according to the other representation.

In lattice simulations, using the variational method, it is possible not only to
reliably separate a given excited state and measure its mass \cite{Mi85} but
also to define and measure a ratio of couplings of different lattice operators
to the given  excited state \cite{GLL}. The two chiral representations
$(0,1)+(1,0)$ and $(1/2,1/2)_b$ form a complete and orthogonal basis (with
respect to the chiral group) for the $\rho$-meson. Consequently using the
variational method we are able to study a mixing of two representations in the
excited $\rho$-meson and see whether or not a given excited $\rho$-meson at low
resolution (infrared) scale couples predominantly to one of the representations and
decouples from the other.

Assuming that the set of interpolating operators $O_i(t)$ is projected to
vanishing spatial momentum, the energies of the states $E^{(n)}$ and the
coefficients giving the overlap of  operators with the physical state,
$
a_i^{(n)}=\langle 0| O_i|n\rangle\;,
$
can be extracted from the cross-correlation matrix
\begin{equation}\label{corr_inf}
C(t)_{ij}=\langle O_i(t)O_j^\dagger(0)\rangle=\sum_n a_i^{(n)} a_j^{(n)*}
\mathrm{e}^{-E^{(n)} t}\;.
\end{equation}

It can be shown \cite{Mi85} (for a recent discussion see
\cite{BuHaLa08,BlDeHi09}) that the generalized eigenvalue problem
\begin{equation}\label{gev_1}
\widehat C(t)_{ij} u_j^{(n)} =\lambda^{(n)}(t,t_0)\widehat C(t_0)_{ij} u_j^{(n)}
\end{equation}
allows to recover the correct eigenvalues and eigenvectors within some
approximation. At $t=t_0$ all eigenvalues  are 1 and the eigenvectors are
arbitrary. Energies of the subsequent states can be extracted from the leading
exponential decays of each eigenvalue.

Ratios of couplings of the different  operators to the physical states can be
obtained as (cf., \cite{GLL})
\begin{equation}\label{ratio_op_comp}
\frac{C(t)_{ij} u_j^{(n)}}{C(t)_{kj} u_j^{(n)}}=\frac{a_i^{(n)}}{a_k^{(n)}}\;.
\end{equation}

A decomposition of a hadron depends on the resolution scale, i.e., what we see
in our microscope, depends on its resolution. If one uses point-like lattice
interpolators then the resolution scale $1/R$ is determined by the lattice
spacing $a$. We are interested to study a decomposition of a hadron at a very
low resolution scale, determined by the hadron size. For that we
cannot use a large $a$ because a proper matching with the ultraviolet
(continuum) behavior of QCD will be lost. Given a fixed, reasonably small, value
for $a$, a small resolution scale $1/R$ can be achieved by the gauge-invariant
smearing of the point-like interpolators \cite{GLL}. We use the interpolator
smeared over the size $R$ in physical units such that $R/a \gg 1$, so even in
the continuum limit $a \rightarrow 0$ we probe the hadron structure at the scale
fixed by $R$. Changing the smearing size $R$, we can study the hadron content at
different resolution scales of the continuum theory at $a \rightarrow 0$. For
this purpose we use Jacobi smearing \cite{GuskenBest},  which provides a
Gaussian shape of   interpolators in spatial directions of size $R$ at both
source and sink. Such definition of the resolution scale is similar to the
experimental definition where a resolution is determined by the momentum
transfer in spatial directions.

\paragraph{3.~Lattice simulation and results.}

\begin{figure}[t]
\begin{center}
\includegraphics[width=85mm,clip]{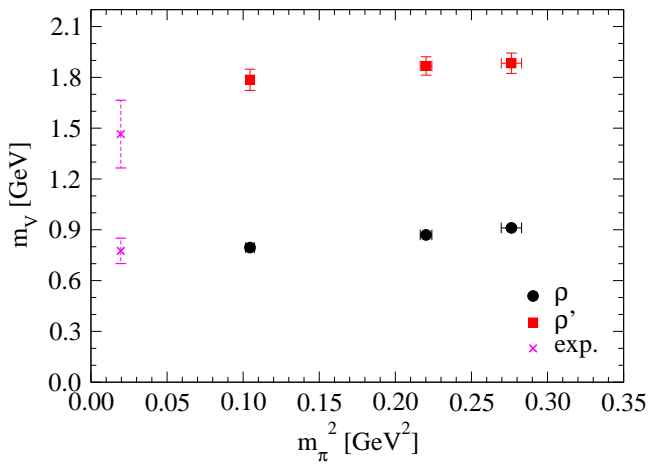}
\caption{\label{fig:masses}
The vector meson mass $m_V$ is plotted against $m_\pi^2$ for all three sets.
Black circles represent the ground state, $\rho$, and red squares represent the
first excitation, $\rho'$. The experimental values are depicted
as magenta crosses with decay width indicated (color online).}
\end{center}
\vspace*{-10pt}
\end{figure}

\begin{table}[tb]
\caption{\label{tab_data}
Specification of the data used here; for the gauge coupling only the leading
value $\beta_{LW}$ is given, $m_0$ denotes the bare mass parameter of the CI
action. Further details on the action, the simulation and the determination
of the lattice spacing and the
$\pi$- and $\rho$-masses are found in \cite{GaHaLa08,Engel:2010my}.}
\begin{center}
\begin{tabular}{ccccccc}
\hline
\hline
Set&  $\beta_{LW}$ &  $a\,m_0$ & \#{conf} & $a$ [fm] & $m_\pi$ [MeV] & $m_\rho$ [MeV]\\
\hline
A & 4.70 & -0.050 & 200 & 0.1507(17) & 526(7) & 911(11) \\
B & 4.65 & -0.060 & 300 & 0.1500(12) & 469(5) & 870(10) \\
C & 4.58 & -0.077 & 300 & 0.1440(12) & 323(5) & 795(15) \\
\hline
\hline
\end{tabular}
\end{center}
\vspace*{-12pt}
\end{table}
\begin{figure}[t]
\begin{center}
\includegraphics[width=85mm,clip]{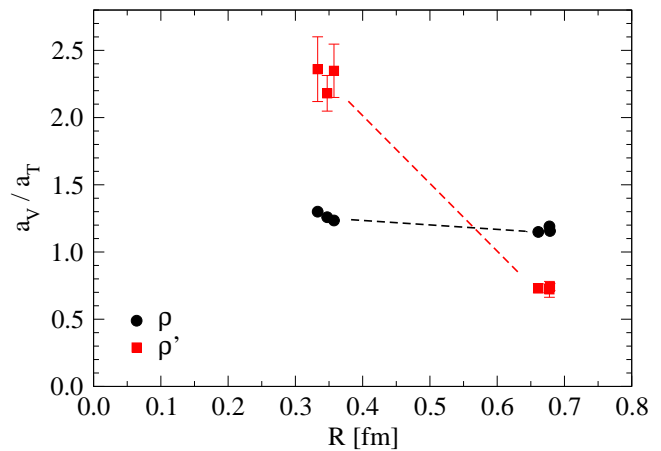}
\caption{\label{fig:ratio}
The ratio $a_V/a_T$ is plotted against the smearing width $R$ for all three data
sets. Black circles represent the ground state and red squares the first
excitation. Broken lines are drawn only to guide the eye (color online).}
\end{center}
\vspace*{-12pt}
\end{figure}

In our excited hadron spectroscopy program, first for quenched configurations
\cite{BuGaGl04a,BuGaGl06,GaGlLa08} and recently for dynamical fermions
\cite{GaHaLa08,Engel:2010my}  the L\"uscher-Weisz gauge action is used
\cite{LuWe85}. For the fermions  the  Chirally Improved (CI) Dirac operator is
adopted, which has better chiral properties than the Wilson Dirac operator
\cite{Ga01a}. 

In this study of the chiral decomposition of the excited  $\rho$-meson we use
three sets of dynamical configurations with two mass-degenerate light sea quarks
on  lattices of size $16^3\times 32$ (see Tab.~\ref{tab_data}).

The following set of four interpolators is used,
\begin{eqnarray}
&O_1=\overline u_n \gamma^i d_n\;,\;\;
&O_2=\overline u_w \gamma^i d_w\;,\\
&O_3=\overline u_n \gamma^t \gamma^i  d_n\;,\;\;
&O_4=\overline u_w \gamma^t \gamma^i  d_w\;.
\end{eqnarray}
$\gamma^i$ is one of the spatial Dirac matrices, $\gamma^t$ is the
$\gamma$-matrix in (Euclidean) time direction, and the subscripts $n$ and $w$
(for narrow and wide)  denote the two smearing widths, $R\approx$ 0.34 fm  and
0.67 fm, respectively. This bunch of operators allows one to extract both the
ground and the first excited state of the $\rho$-meson, see
Fig.~\ref{fig:masses}.

In Fig.~\ref{fig:ratio} we show the $R$-dependence of the ratio $a_V/a_T$ both
for the ground state $\rho$-meson and its first excited state. This ratio
of the vector to pseudotensor coupling gives us the information on the
chiral decomposition of both states in terms of the representations
$(0,1)+(1,0)$ and $(1/2,1/2)_b$. We observe that this ratio for the ground state
at the smallest possible resolution scale (largest smearing radius) $R=0.67$ fm approaches a value close
to $1.2$. This implies a strong mixture of both representations in the
$\rho$-meson wave function.  Using a unitary transformation from the chiral
basis to the $^{2S+1}L_J$ basis \cite{GN},
\begin{eqnarray}\label{chiral_basis}
|{}^3S_1\rangle &=& \sqrt{\frac{2}{3}}\, |(0,1)+(1,0); 1^{--}\rangle + \nonumber\\
                & & \qquad\qquad \sqrt{\frac{1}{3}}\, |(1/2,1/2)_b;1^{--}\rangle\;,\nonumber\\
|{}^3D_1\rangle &=& \sqrt{\frac{1}{3}}\, |(0,1)+(1,0); 1^{--}\rangle -\nonumber\\
                & & \qquad\qquad \sqrt{\frac{2}{3}}\, |(1/2,1/2)_b; 1^{--}\rangle\;,
\end{eqnarray}
one obtains that the ground state $\rho$ is a $^3S_1$ state with a tiny
admixture of the $^3D_1$ wave, $0.997|{}^3S_1\rangle - 0.073 |{}^3D_1\rangle$.
This implies that at a resolution fixed by the $\rho$-size
 the $\rho(770)$ is approximately a $^3S_1$
state  in agreement with the quark model.

Our main result in the present report is that the chiral decomposition of the
first excited state, $\rho' = \rho(1450)$, and its scale dependence is  very
different.  In this case at large $R$ a leading contribution comes from the
$(1/2,1/2)_b$ representation. Given only two different $R$ values  we cannot
reliably extrapolate to the scale of 1 fm, as suggested by the size of excited 
rho-meson. It is clearly seen, however,
that such a ratio at  $R  \sim 1$ fm is very small; it can be both positive and
negative in sign.

This behavior has a clear interpretation. In the deep ultraviolet one expects
from the conformal symmetry of QCD that the pseudotensor interpolator decouples
from the $\rho$-mesons. Consequently towards small $R$ the ratio $a_V/a_T$ must
increase. At large $R$ the ratio determines a degree of the chiral symmetry
breaking in the infrared region, where mass is generated. One observes that such
a breaking for the $\rho(1450)$ is  insignificant, in contrast to the
$\rho(770)$. Since the chiral decomposition of the $\rho'$ state is dominated by
one of the chiral representations, it indicates a smooth onset of effective
chiral restoration. Given that this leading representation of $\rho'$ is
$(1/2,1/2)_b$ one predicts that in the same energy region there must exist a
$h_1$ (and not a $a_1$) meson. Inspecting the   Particle Data Group
\cite{PDG} one finds that there is indeed the state $h_1(1380)$ 
and no $a_1$ in the same energy region.

One naturally arrives at the following identification for $\rho$-mesons. Chiral
symmetry is strongly broken in the $\rho(770)$, since it is a strong mixture of
$(0,1)+(1,0)$ and $(1/2,1/2)_b$, and consequently its ``would-be chiral
partners" have a much larger mass: $a_1(1260)$ and  $h_1(1170)$. All these
lowest-lying states  $\rho(770)$, $a_1(1260)$ and  $h_1(1170)$ cannot be
assigned to any chiral representation. In the first excited state, $\rho(1450)$,
a contribution of $(1/2,1/2)_b$ is much bigger than a contribution of the other
representation, and consequently its approximate chiral partner is the
$h_1(1380)$. Then, the next excited $\rho$-meson, $\rho(1700)$, is most probably
dominated by $(0,1)+(1,0)$, which is supported by existence of a nearby $a_1$
state,  $a_1(1640)$. Note that there is no room for this $a_1(1640)$ meson
within the quark model \cite{IS,PDG}.

While we do not know the precise value of the ratio $a_V/a_T$ for the
$\rho(1450)$ at  $R \sim 1$ fm, it is indicative that this ratio is very small. Then we
can qualitatively estimate the angular momentum content of $\rho(1450)$ in the
infrared. Assuming a vanishing ratio, one obtains the following partial wave
content, $\sqrt {1/3} |{}^3S_1\rangle - \sqrt {2/3} |{}^3D_1\rangle$. A
significant contribution of the $^3D_1$ wave is obvious. A possible variation of
the ratio at large $R$ changes slightly numbers for the partial wave
decomposition, but does not change the qualitative result. This result is not
consistent with $\rho'$ to be a  radial excitation of the ground state
$\rho$-meson, i.e., a $^3S_1$ state, as predicted by the quark model.

\begin{acknowledgments}
We thank G.~Engel and C.~Gattringer for discussions. L.Ya.G.~and
M.L.~acknowledge support of the Fonds zur F\"orderung der Wissenschaflichen
Forschung (P21970-N16) and (DK W1203-N08), respectively. The calculations have
been performed on the SGI Altix 4700 of the Leibniz-Rechenzentrum Munich and on
local clusters at ZID at the University of Graz.
\end{acknowledgments}

\end{document}